\documentclass[twocolumn]{aastex62}

\def\simgt{\ {\raise-.5ex\hbox{$\buildrel>\over\sim$}}\ }
\def\simlt{\ {\raise-.5ex\hbox{$\buildrel<\over\sim$}}\ }
\def\I{\'\i}
\def\cd{d$^{-1}$\,}
\def\kms{km\,s$^{-1}$\,}

\received{January 7, 2019}
\revised{January 1, 2019}
\accepted{\today}
\submitjournal{ApJL}

\shorttitle{Asteroseismology of HN~Aqr with {\it TESS}}
\shortauthors{Handler et al.}

\begin{document}

\title{Asteroseismology of massive stars with the {\it TESS} mission: the 
runaway $\beta$~Cep pulsator PHL~346 = HN~Aqr}

\correspondingauthor{Gerald Handler}
\email{gerald@camk.edu.pl}

\author[0000-0001-7756-1568]{Gerald Handler}
\affil{Nicolaus Copernicus Astronomical Center, Bartycka 18, 00-716 Warsaw, Poland} 

\author[0000-0003-2488-6726]{Andrzej Pigulski}
\affiliation{Astronomical Institute Wroc{\l}aw University, ul. Kopernika 11, 51-622 Wroc{\l}aw, Poland}

\author[0000-0001-9704-6408]{Jadwiga Daszy\'nska-Daszkiewicz}
\affiliation{Astronomical Institute Wroc{\l}aw University, ul. Kopernika 11, 51-622 Wroc{\l}aw, Poland}

\author[0000-0002-0465-3725]{Andreas Irrgang}
\affil{Dr. Karl Remeis-Observatory \& ECAP, Astronomical Institute, Friedrich-Alexander University Erlangen-N\"urnberg (FAU)\\ Sternwartstr. 7, 96049 Bamberg, Germany}

\author[0000-0003-4586-0832]{David Kilkenny}
\affil{Department of Physics \& Astronomy, University of the Western Cape, Private Bag X17, Bellville 7535, South Africa}

\author[0000-0002-0951-2171]{Zhao Guo}
\affil{Center for Exoplanets \& Habitable Worlds, Department of Astronomy \& Astrophysics\\ Eberly College of Science, The Pennsylvania State University, 525 Davey Lab, University Park, PA 16802, USA}

\author[0000-0001-5263-9998]{Norbert Przybilla}
\affil{Institut f\"ur Astro- und Teilchenphysik, Universit\"at Innsbruck, Technikerstr. 25/8, 6020 Innsbruck, Austria}

\author[0000-0002-9036-7476]{Filiz Kahraman Ali\c{c}avu\c{s}}
\affil{Nicolaus Copernicus Astronomical Center, Bartycka 18, 00-716 Warsaw, Poland}
\affil{\c{C}anakkale Onsekiz Mart University, Faculty of Sciences and Arts, Physics Department, 17100 \c{C}anakkale, Turkey}

\author[0000-0003-3627-2561]{Thomas Kallinger}
\affil{Institute for Astrophysics, University of Vienna, T\"urkenschanzstrasse 17, 1180 Vienna, Austria}

\author[0000-0003-0139-6951]{Javier Pascual-Granado}
\affil{Instituto de Astrof\I sica de Andaluc\I a (IAA-CSIC), Glorieta de Astronom\I a s\textbackslash n, E-18008 Granada, Spain}

\author[0000-0001-7290-5800]{Ewa Niemczura}
\affil{Astronomical Institute Wroc{\l}aw University, ul. Kopernika 11, 51-622 Wroc{\l}aw, Poland}

\author[0000-0002-5819-3023]{Tomasz R\'o\.za\'nski}
\affil{Astronomical Institute Wroc{\l}aw University, ul. Kopernika 11, 51-622 Wroc{\l}aw, Poland}

\author[0000-0001-7444-5131]{Sowgata Chowdhury}
\affil{Nicolaus Copernicus Astronomical Center, Bartycka 18, 00-716 Warsaw, Poland}

\author[0000-0002-1988-143X]{Derek L. Buzasi}
\affil{Dept. of Chemistry \& Physics, Florida Gulf Coast University,
10501 FGCU Blvd. S., Fort Myers, FL 33965, USA}
  
\author[0000-0003-0238-8435]{Giovanni M. Mirouh}
\affil{Astrophysics Research Group, Faculty of Engineering and Physical Sciences, University of Surrey, Guildford GU2 7XH, UK}


\author[0000-0001-7402-3852]{Dominic M. Bowman}
\affil{Instituut voor Sterrenkunde, KU Leuven, Celestijnenlaan 200D, 3001 Leuven, Belgium}

\author[0000-0002-3054-4135]{Cole Johnston}
\affil{Instituut voor Sterrenkunde, KU Leuven, Celestijnenlaan 200D, 3001 Leuven, Belgium}

\author[0000-0002-7950-0061]{May G. Pedersen}
\affil{Instituut voor Sterrenkunde, KU Leuven, Celestijnenlaan 200D, 3001 Leuven, Belgium}

\author[0000-0003-1168-3524]{Sergio Sim\'on-D\I az}
\affil{Instituto de Astrof\I s\I ca de Canarias, E-38200 La Laguna, Tenerife, Spain}
\affil{Departamento de Astrof\I s\I ca, Universidad de La Laguna, E-38205 La Laguna, Tenerife, Spain}

\author[0000-0003-4372-0588]{Ehsan Moravveji}
\affil{Instituut voor Sterrenkunde, KU Leuven, Celestijnenlaan 200D, 3001 Leuven, Belgium}

\author[0000-0002-8855-3923]{Kosmas Gazeas}
\affil{Section of Astrophysics, Astronomy and Mechanics, Department of Physics, National and Kapodistrian University of Athens, Zografos GR-15784, Athens, Greece}

\author[0000-0001-5419-2042]{Peter De Cat}
\affil{Royal Observatory of Belgium, Ringlaan 3, B-1180 Brussel, Belgium}

\author[0000-0001-6763-6562]{Roland K. Vanderspek}
\affil{Department of Physics, and Kavli Institute for Astrophysics and Space Research, Massachusetts Institute of Technology, Cambridge, MA 02139, USA}

\author[0000-0003-2058-6662]{George R. Ricker}
\affil{Department of Physics, and Kavli Institute for Astrophysics and Space Research, Massachusetts Institute of Technology, Cambridge, MA 02139, USA}



\begin{abstract}

We report an analysis of the first known $\beta$~Cep pulsator 
observed by the {\it TESS} mission, the runaway star PHL~346 = HN~Aqr. The 
star, previously known as a singly-periodic pulsator, has at least 34
oscillation modes excited, 12 of those in the g-mode domain and 22 p~modes.
Analysis of archival data implies that the amplitude and frequency of the 
dominant mode and the stellar radial velocity were variable over time. A 
binary nature would be inconsistent with 
the inferred ejection velocity from the Galactic disc of 420\,\kms, 
which is too large to be survivable by a runaway binary system. A 
kinematic analysis of the star results in an age constraint 
($23\pm1$\,Myr) that can be imposed on asteroseismic modelling and that
can be used to remove degeneracies in the modelling process.
Our attempts to match the excitation of the observed frequency 
spectrum resulted in pulsation models that were too young. Hence, 
asteroseismic studies of runaway pulsators can become vital not only in tracing 
the evolutionary history of such objects, but to understand the interior
structure of massive stars in general. {\it TESS} is now opening up these
stars for detailed asteroseismic investigation.

\end{abstract}



\section{Introduction} \label{sec:intro}

The Transiting Exoplanet Survey Satellite ({\it TESS}) is a NASA mission whose 
primary objective is to discover hundreds of transiting planets smaller 
than Neptune with host stars bright enough for spectroscopic follow-up 
to measure planetary masses and atmospheric compositions \citep{ricker}. 
{\it TESS} has commenced its almost-all-sky survey of bright stars ($4< I_c<13$)
in a wide red-bandpass filter. In the first two years of operation,
precision time series photometry is obtained for 200\,000 pre-selected
targets with 2-min cadence, for about 30 million stars every 30 minutes.


The characterization of extrasolar planets requires information about 
their host stars. One of the methods that yields this information is 
asteroseismology \citep[e.g.,][]{LHSC18}, the study of stellar interiors 
by utilizing their pulsations as seismic waves. Asteroseismology comes 
at no additional cost to planet-search photometric missions as the 
observational technique is essentially the same: high-accuracy 
time-resolved photometry. Consequently, asteroseismology and exoplanet 
space missions often are combined \citep[e.g.,][]{M06,G10}, just like 
{\it TESS} does.

Because {\it TESS} will be the first precision photometry mission that surveys 
almost the whole sky, some types of stars that were not prime targets 
for searches for extrasolar planets will now be observed in large 
amounts. In particular, the asteroseismic potential of hot massive stars 
does not appear to have been fully exploited yet. \citet{P19} give a 
first overview of what {\it TESS} can do for OB star astrophysics. In this 
Letter, we report a study of the first known $\beta$~Cep pulsator 
observed with the {\it TESS} mission, PHL~346 = HN~Aqr = TIC~69925250.

\subsection{PHL~346 = HN~Aqr} \label{subsec:phl346}

PHL~346 is a star of $V=11.44$ located at a Galactic latitude of $b 
\approx58^o$. \citet{KHB77} classified it as spectral type B1, and 
\citet{K86} reported a surface gravity consistent with an evolved 
main-sequence evolutionary status and Pop.~I metal abundances. Given the 
radial velocity they measured ($+66\pm10$\,\kms), \citet{K86} had to 
conclude that PHL~346 would not have had enough time to attain such a 
high Galactic latitude within its lifetime and thus must have been 
formed far from the Galactic plane. This puzzling result was amended by 
\citet{RHM01} who, based on a new spectroscopic analysis and the first 
proper motion measurement of PHL~346, reconciled the stellar flight time 
with its lifetime, meaning the star could have been formed in and 
ejected from the Galactic plane.

PHL~346 was the subject of several studies that resulted in 
determinations of its effective temperature and surface gravity, summarized
in Table~\ref{tab:parameters} and available online only.
$\beta$~Cep-type pulsations of PHL~346 were discovered by \citet{WR88} 
and confirmed by \citet{KW90}. The star was also observed during the 
ASAS survey \citep{PP08} and by \citet{HS08}. All these authors detected 
the same single oscillation frequency near 6.566\,\cd. \citet{HWS94} and 
\citet{C94} suggested this pulsation is due to a nonradial $l=1$ mode. 
On the other hand, \citet{HS08} derived $l=2$ or 4 for this oscillation, 
but noted a possible problem with the $U$ filter data their 
identification critically hinged upon.

\section{Observations} \label{subsec:tess}

\begin{figure}[htb!]
\includegraphics[width=\columnwidth]{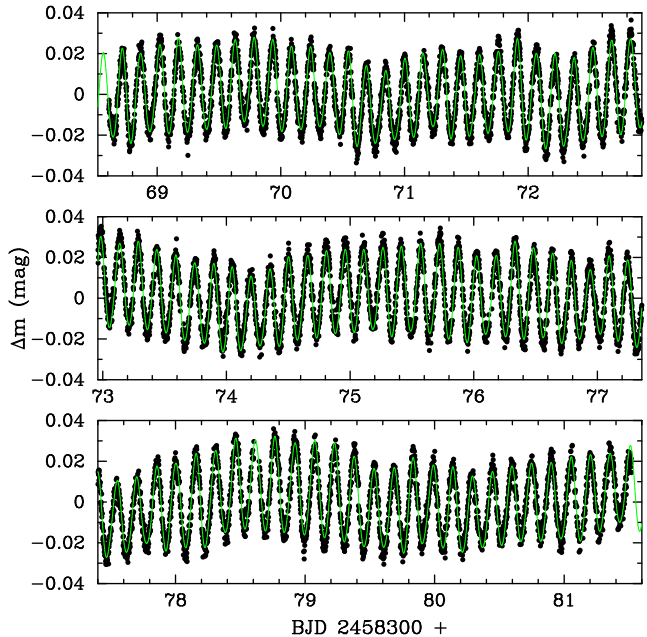}
\caption{Light curve from 13 days of {\it TESS} observations of HN~Aqr
  (black dots). Overplotted in green is a multifrequency fit to be derived in Sect.~\ref{subsec:periods}.\label{fig:lc}}
\end{figure}

PHL~346 was observed with the {\it TESS} mission in Sector~2, from August~23, 
2018 to September~20, 2018 in 2-min cadence. 18317 brightness 
measurements were secured over a time base of $\Delta T=27.4$\,d, with a 
duty cycle of 92.8\%. 
The photometry was downloaded from 
MAST\footnote{\url{https://mast.stsci.edu/portal/Mashup/Clients/Mast/Portal.html}}, 
and the PDC\_SAP fluxes were converted into magnitude. No further 
manipulations with the data were made. The second part of the light 
curve is shown in Fig. \ref{fig:lc}. Clearly, HN~Aqr is not a 
singly-periodic variable. Variations in the mean light level that
are not singly periodic as well as 
changes in the amplitude of the dominant short-period signal are 
visible.

\section{Analysis}

\subsection{Periodicities in the {\it TESS} light curve} \label{subsec:periods}

Given that HN~Aqr is the first $\beta$~Cep pulsator observed with {\it TESS}, 
given that its light variations are multiperiodic, and given that the 
time base of the observations may not allow all pulsational signals to 
be resolved in frequency, we analyzed the data with various methods. 
Method 1 comprised classical single-frequency power spectrum analysis 
and simultaneous multi-frequency sine-wave fitting with input parameter 
optimization. The sine-wave fits are subtracted from the data and the 
residuals examined for the presence of further periodicities 
(``prewhitening''). The noise level was assessed in apparently 
signal-free frequency regions and the $S/N>4$ criterion by \citet{B93} 
was adopted to evaluate the significance of signal detection.

Method 2 used essentially the same approach, but the background noise in 
the Fourier spectrum was modelled in log-log space following \citet{P17} 
using the amplitude spectrum from 0.036 to 30\,\cd prewhitened with the 
strongest oscillation. Again, the $S/N>4$ criterion was used. Method 3 
applied the MIARMA gap-filling method \citep{PGS15} to improve the 
spectral window function of the data. This modified light curve was then 
frequency analyzed with the SigSpec algorithm \citep{R06} that also 
involves pre-whitening.

Method 4 \citep{KW16} employed a Bayesian algorithm allowing a 
probabilistic assessment of the significance of signal detection. It was 
run twice, once allowing only a single frequency to represent a peak in 
the amplitude spectrum (equivalent to multifrequency sine-wave fitting), 
and the other time allowing close frequency doublets (spacing $\leq 
3/\Delta T=0.11$\,\cd). Finally, Method 5 used a Morley wavelet 
transform \citep{TC98}. Output from this approach reflected the 
amplitude modulation of the short-period pulsations 
(Fig.\,\ref{fig:lc}). In the low-frequency region ($f<2$\,\cd), 
considerable changes in the amplitudes of individual signals were 
visible, in most cases with some degree of regularity indicating 
multifrequency beating.

To reconcile the outcome of the different methods we have compared the 
results of the first four techniques that resulted in lists of 
frequencies and amplitudes. They yielded fairly consistent results for 
strong, well-resolved signals, but diverged in frequency regions where 
densely spaced signals were present. The number of detected frequencies 
strongly depended on the adopted signal detection threshold.

To provide a set of pulsation frequencies that can be reasonably safely 
used for asteroseismic investigations, we only accepted signals detected 
by at least three of the methods independently. The amplitudes of these 
signals had to exceed the noise level $a(\nu)$ (Eq.\,1) by factors of 5 
(independent signals) and 3.5 (combination frequencies), respectively. 
The more conservative S/N threshold was chosen to avoid picking up 
spurious frequencies in a data set of the present size and sampling
\citep{BKP15}; 
the noise level was computed according to Method 2:

\begin{equation} 
  a(\nu)=\frac{a_0}{1+(\frac{\nu}{\nu_0})^\lambda}+P_0,
\end{equation} 

where $P_0=1.4388\times10^{-5}$\,mag is a constant Gaussian noise term, 
$\lambda=2.0$, $a_0=7.76\times10^{-5}$\,mag and $\nu_0=(2\pi\tau)^{-1}, 
\tau=0.076$\,d are the amplitude and characteristic frequency describing 
the red noise, respectively.

\begin{figure}[htb!]
\includegraphics[width=\columnwidth]{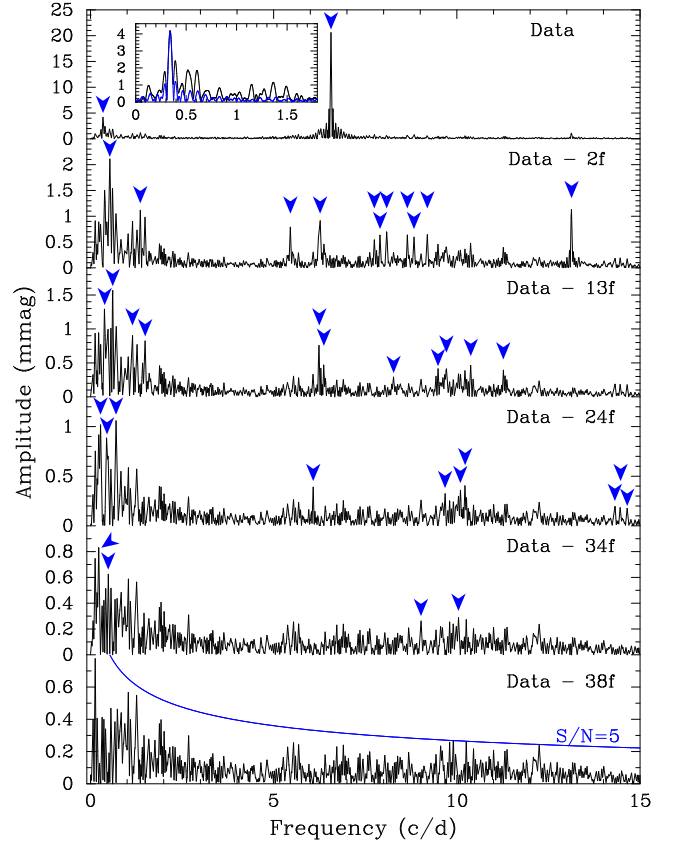}
\caption{Fourier spectra of the {\it TESS} observations of HN~Aqr. The blue 
arrows denote frequencies that are prewhitened in the panel below and 
correspond to the 38 signals listed in Table \ref{tab:frequencies}. The
inset in the uppermost panel shows a comparison of the spectral window
(blue) and amplitude spectrum (black) in the low-frequency domain.
\label{fig:ft}}
\end{figure}

The resulting list of frequencies is given in 
Table\,\ref{tab:frequencies}. The frequency values and error estimates 
were adopted from the Bayesian method as it does not rely on 
prewhitening. The error bars correspond well to the least-squares errors 
\citep{MO99} for well-separated signals, but take systematic 
uncertainties of closely-spaced frequencies into account. As it is not 
yet well known how much data processing affects the low-frequency 
domain, periods longer than 4\,d should be treated with caution. Some 
steps of prewhitening of the {\it TESS} data are illustrated in 
Fig.\,\ref{fig:ft}. Our multifrequency fit leaves a residual scatter of 
3.4 mmag per point (617 ppm/hr) containing residual systematic (presumably
mostly stellar) variability contributing some 15\% to the total scatter.

\startlongtable
\begin{deluxetable}{lccr}
\tablecaption{Multifrequency solution for our {\it TESS} photometry of HN~Aqr.
Error estimates for the independent frequencies are given in braces
in units of the last significant digit; the errors on the amplitudes
are about $\pm0.03$\,mmag.\label{tab:frequencies}}
\tablehead{
\colhead{ID} & \colhead{Freq.} & \colhead{Ampl.} & \colhead{S/N} \\
\colhead{} & \colhead{(\cd)} & \colhead{(mmag)} & \colhead{}
}
\startdata
$f_{1}$& 0.2241(8)& 0.93 & 5.1 \\ 
$f_{2}$& 0.2731(7)& 0.88 & 5.0 \\ 
$f_{3}$& 0.3421(2)& 4.00 & 24.2 \\ 
$f_{4}$& 0.3881(5)& 1.55 & 9.7 \\ 
$f_{5}$& 0.4522(10)& 0.84 & 5.5 \\ 
$f_{6}$& 0.4844(17)& 0.75 & 5.0 \\ 
$f_{7}$& 0.5260(4)& 2.06 & 14.0 \\ 
$f_{8}$& 0.6058(4)& 1.67 & 11.8 \\ 
$f_{9}$ \tablenotemark{a}& 0.697(1)& 1.09 & 8.0 \\ 
$f_{10}$& 1.1461(7)& 0.89 & 7.7 \\ 
$f_{11}$& 1.3589(6)& 1.12 & 10.3 \\ 
$f_{12}$& 1.4919(7)& 0.87 & 8.2 \\ 
$f_{13}$& 5.4562(8)& 0.79 & 12.4 \\ 
$f_{14}$& 6.079(1)& 0.39 & 6.4 \\ 
$f_{15}$& 6.245(2)& 0.83 & 13.8 \\ 
$f_{16}$& 6.267(1)& 1.06 & 17.6 \\ 
$f_{17}$& 6.371(2)& 0.38 & 6.4 \\ 
$f_{18}$ \tablenotemark{b}& 6.5653(5)& 20.65 & 349.4 \\ 
$f_{19}$& 7.747(1)& 0.60 & 10.9 \\ 
$f_{20}$& 7.900(1)& 0.58 & 10.7 \\ 
$f_{21}$& 8.084(1)& 0.69 & 12.7 \\ 
$f_{22}$& 8.273(3)& 0.29 & 5.4 \\ 
$f_{23}$& 8.649(1)& 0.64 & 12.1 \\ 
$f_{24}$& 8.829(1)& 0.54 & 10.5 \\ 
$f_{25}$& 9.021(2)& 0.26 & 5.0 \\ 
$f_{26}$& 9.194(1)& 0.61 & 11.9 \\ 
$f_{27}$& 9.486(2)& 0.45 & 8.9 \\ 
$f_{28}$& 9.677(2)& 0.38 & 7.6 \\ 
$f_{29}$& 9.714(2)& 0.36 & 7.3 \\ 
$f_{30}$& 10.043(2)& 0.37 & 7.6 \\ 
$f_{31}$& 10.095(3)& 0.29 & 6.0 \\ 
$f_{32}$& 10.226(2)& 0.34 & 7.0 \\ 
$f_{33}$& 10.380(2)& 0.44 & 9.0 \\ 
$f_{34}$& 11.267(2)& 0.37 & 8.0 \\ 
2$f_{18}$& 13.1306(7)& 1.13 & 26.1 \\ 
$f_{18}+f_{19}$& 14.312(1)& 0.19 & 4.5 \\ 
$f_{18}+f_{20}$& 14.468(1)& 0.15 & 3.6 \\ 
$f_{18}+f_{21}$& 14.649(1)& 0.18 & 4.3 \\ 
\enddata
\tablenotetext{a}{Likely a close doublet (0.679/0.715\,\cd).}
\tablenotetext{b}{Possibly a close doublet (6.555/6.5666\,\cd).}
\end{deluxetable}

The low-frequency domain of detected signals extends up to 1.49\,\cd. 
As the large number of frequencies cannot be explained by effects of
binarity or rotation most, if not all, of these must be due to pulsation.
There is 
no evidence of rotational or binary-induced modulation, or of a regular 
period spacing. A re-occurring frequency spacing is present within the 
signals in the p-mode domain ($\Delta f\approx0.187$\,\cd), which may be 
due to rotational splitting. No multiplets were identified with 
confidence; the best candidate consists of frequencies $f_{22}-f_{25}$ 
which could be part of an $l=2$ quintuplet with the $m=0$ mode missing.

\subsection{Stability of the period and amplitude}

Thirty years have passed since pulsation in PHL~346 was discovered. Thus 
the long-term stability of the period and amplitude of the dominant 
pulsation mode can be studied. We gathered all available time-series 
photometry of the star \citep{WR88,KW90,HS08,PP08}. Additionally, we 
include some previously unpublished observations from 1989 and 1990 by 
one of us (DK), IUE spectrophotometry \citep{D96}, as well as NSVS
\citep{W04}, and ASAS-SN \citep{S14} data. Each of the individual data
sets used spans at least 13\,d.

\begin{figure}
\includegraphics[width=\columnwidth]{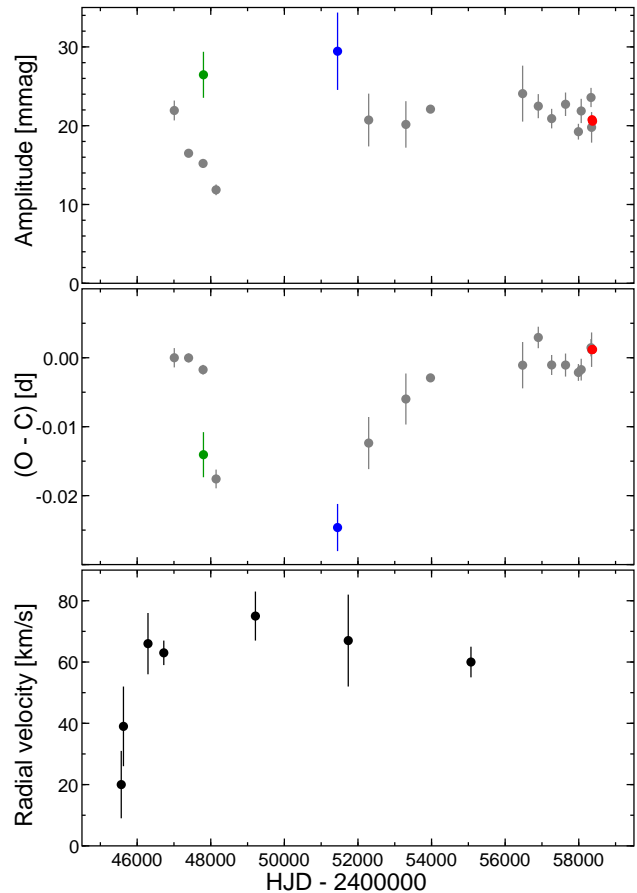}
\caption{Top: Amplitude of the dominant mode in HN~Aqr. The values 
derived from IUE, NSVS, and {\it TESS} photometry are shown with green, blue, 
and red dots, respectively, the remaining ones with grey dots. Middle: 
The O\,--\,C diagram for the times of maximum light of this mode 
calculated using the ephemeris: $T_{\rm 
max}=2447008.89183+0.15231625\times E$, where $E$ is the number of 
pulsation cycles elapsed since the initial epoch. Bottom: Radial 
velocities of HN~Aqr. The x-axis starts in 1982 and spans roughly 40 years.}
\label{fig:o-c}
\end{figure}

The resulting O\,--\,C diagram and amplitudes over time are shown in 
Fig.\,\ref{fig:o-c}. Taking into account passband differences between 
the various individual data sets, the amplitude of the dominant mode 
remained relatively stable over 30 years at a level of about 22~mmag and 
dropped only in the late 1980s and early 1990s, down to 
11.9\,$\pm$\,0.7~mmag in 1990. (As expected, the amplitude in the UV was 
higher than in the visual.) During this drop, the O\,--\,C diagram shows 
a significant period change: the 1990 value of O\,--\,C is $\approx$ 25 
minutes off the ephemeris. We checked whether the shape of the O\,--\,C 
diagram can be explained in terms of the light-time effect in a binary 
system, with no satisfactory solution. In any case, such an orbit needs 
to be highly eccentric ($e\simgt 0.7$) with periastron passage in the 
mid 1990's, when unfortunately a gap of almost a decade in the 
photometric data occurs. This would be also the time when large 
radial-velocity (RV) changes would occur. We therefore gathered all 
available radial-velocity data for HN~Aqr 
\citep{K86,KM89,H96,RHM01,L02}, and plot them in the lower panel of 
Fig.\,\ref{fig:o-c}. Although there is no large change of RV at the 
predicted time of the periastron passage, the first two RVs 
\citep{KM89}, are significantly different.

Assuming a period of 30~yr for a hypothetical binary orbit of HN~Aqr
and $M=10$\,M$_{\odot}$ for the primary,
a companion of $M_2\sin i=1.15$\,M$_{\odot}$ would be able to explain the
0.013\,d semi-amplitude of the (O--C) variations. However, such a companion
would cause an orbital RV semi-amplitude of only 2.2\,\kms, more than
an order of magnitude less than the observed RV change. To summarize,
in view of the large uncertainties of RVs, the presence of 
pulsations that increase the scatter of RVs, the large gap in 
photometric data which causes cycle count ambiguities in the 1990s,
the inconsistency of the hypothetical light time effect and the radial
velocity change, the results as to the binarity of the star are
inconclusive. What one can conclude, however, is that the observed
pulsational (O--C) variations and RV changes cannot be caused by
binarity alone.


 
\subsection{Spectroscopy and kinematics} \label{subsec:spectro}

High-resolution ($R\approx35\,000$) spectra of HN~Aqr were taken with 
the UVES spectrograph \citep{D00} attached to the VLT-UT2 at the 
European Southern Observatory (ESO) on the night of August~27, 2009. 
Integration times of 1150\,s at both the blue and the red arm led to S/N 
ratios of about $150-190$ in the range of $304-669$\,nm, and around 70 
up to 1043\,nm. The parameters derived from these spectra using two
analysis strategies \citep{I14, HL17}
are listed in online Table~\ref{tab:parameters}.

Summarizing all these results and using bolometric corrections from
\citet{F96} suggests that a range 
of basic parameters $T_{\rm eff}=22300\pm900$\,K, log $g=3.75\pm0.15$, 
$M_\mathrm{bol}=-5.2\pm0.3$ and $[M/H]=0.2\pm0.1$ should contain the 
parameter space in which a seismic model for HN~Aqr ought to be located. 
Furthermore, a projected rotational velocity $v \sin i=30\pm5$\,\kms
can be adopted. The Gaia DR2 parallax $\pi=0.10\pm0.08$ mas 
for HN~Aqr \citep{GC16, GC18, L18} cannot be used to better constrain 
the stellar luminosity, but is consistent with the results above.

The increased metallicity agrees very well with a kinematic 
investigation based on the radial velocity, Gaia DR2 proper motions, and 
a derived spectrophotometric distance of $6.7\pm0.8$\,kpc (99\% 
confidence interval; assuming a mass of $9.5\pm0.4\,M_\odot$). It 
suggests that the star stems from the inner part of the Galaxy 
(Galactocentric radius at disk intersection $2.3\pm0.4$ kpc with 68\% 
confidence interval). After taking the Galactic abundance gradients 
into consideration, the abundance pattern appears to be perfectly normal 
except for an underabundance of $\sim0.3$ dex in carbon.
From the calculation of stellar trajectories \citep[for details, 
see][]{I13}, we infer a flight time to the Galactic disk of $23\pm1$ Myr 
(68\% confidence interval).

\subsection{Pulsational modeling} \label{subsec:modeling}

\begin{figure}
\includegraphics[width=\columnwidth]{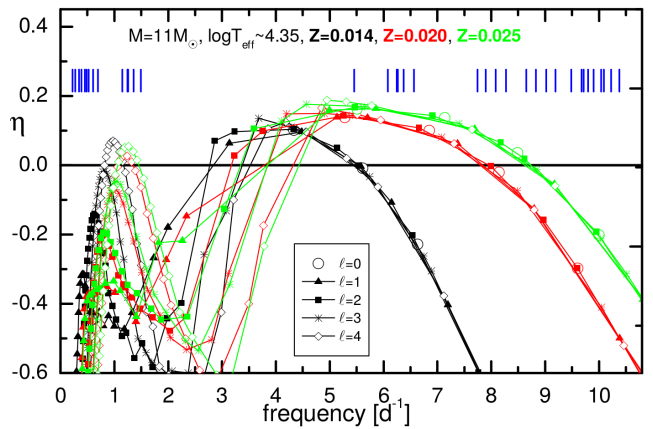}
\caption{Effect of metallicity on pulsational driving of modes with
  $0\leq l\leq 4, m=0$ for a 11\,$M_{\sun}$ model of HN~Aqr. Theoretical
  modes with a stability parameter $\eta>0$ are driven; the vertical
  blue bars denote the observed independent frequencies. An initial
  hydrogen abundance $X=0.71$ \citep{NP12}, OPAL opacities, the \cite{A09}
element mixture and no convective core overshooting were used.}
\label{fig:model}
\end{figure}

We follow \citet{DD17} to model the pulsation spectrum of HN~Aqr 
preliminarily. The basic stellar parameters determined earlier are 
represented best by models of 11\,$M_{\sun}$ that we take for the 
purpose of example. We attempt to reproduce the frequency domains of the 
observed p and g-modes (Sect. \ref{subsec:periods}) and search for 
required modifications of the stellar models. Given the low projected 
rotational velocity of the star, and that the suspected first-order
p-mode splitting from Sect.\,\ref{subsec:periods} would suggest
$v_{\rm rot} \approx 64$\,\kms, the observed frequency ranges should 
not differ much from those in the stellar frame of rest. In addition to 
the frequencies, we consider the age constraint of $23\pm1$ Myr from the 
kinematic analysis. The most important influence on mode stability is 
provided by the overall metallicity, as shown in Fig.\,\ref{fig:model}.

However, increasing $Z$ alone, consistent with the spectroscopically
determined abundances (Table~\ref{tab:parameters}),
which gives a better match to the p-mode
domain, is insufficient to reproduce the observed g-mode region. Also,
increasing $Z$ requires more massive models that evolve too rapidly: 
the model with $Z=0.025$ has an age of 13.5 Myr and a main sequence 
lifetime of 18.0 Myr. Possible ways out of this problem would be the 
inclusion of convective core overshooting prolonging the main sequence 
lifetime (a 11\,$M_{\sun}$, $Z=0.025$, $\alpha_{\mathrm ov}=0.2$, $\log 
T_{\rm eff}=4.35$ model has an age of 15.5 Myr), increasing the He 
abundance, leading to models of lower mass, and modifications of the 
input opacities, in particular near $\log T=5.46$, corresponding to an 
enhancement of the nickel opacity \citep[see, e.g.][]{DD17}.

\section{Summary and conclusions}

{\it TESS} photometry of the pulsating runaway star HN~Aqr provided the 
detection of 38 frequencies of variability (34 independent modes within 
them) with evidence for more. The star has rich p- and g-mode pulsation 
spectra. Some of the oscillation frequencies are formally unresolved 
during the 27.4-d time base of the observations, a problem expected to 
affect the analysis of other $\beta$~Cep and related types of pulsating 
stars as well. Hence we applied five different frequency analysis 
techniques whose combination resulted in a reliable solution.

The signals in the low-frequency domain are dominated by g-mode 
pulsation. The frequency spectrum is denser at the low-frequency end, 
which is expected for opacity-driven g modes, but the occurrence of 
internal gravity waves \citep[e.g.,][]{AR15,B19} (or residual instrumental 
effects) may also be suspected. The p-mode frequency region is 
surprisingly wide and spans some five radial overtones (cf. 
Fig.\,\ref{fig:model}); we detected 22 independent frequencies in this 
domain.


An analysis of archival data showed that the frequency and amplitude of 
the dominant mode, as well as the radial velocity of the star were not 
stable over the last 30 years. We could speculate about the presence of 
a binary companion, but this would be rather surprising because such 
systems should not survive the ejection of the star from the Galactic 
disc \citep{PS12}, and capture of a companion by a fast-moving runaway 
star is unfeasible. Furthermore, our Bayesian frequency analysis 
provided evidence that the dominant pulsation frequency of HN~Aqr may be 
a close doublet, which would provide an alternative interpretation for 
its amplitude and frequency variations. Therefore, the star should be 
included in a long-term spectroscopic and photometric observing program.

Pulsational modeling shows that the observed pulsation spectrum cannot 
be reproduced by increasing the metallicity only; an increase in the 
opacities in certain stellar interior regions is required. 
Interestingly, and perhaps most importantly, the nature of HN~Aqr as a
runaway star provides 
additional constraints on the modeling, as the age of the models must be 
reconciled with the stellar flight time ($23\pm1$\,Myr). Our initial 
attempts in this direction were unsuccessful as we could not obtain 
models older than 18 Myr that would give a reasonable match to the 
observed pulsation spectrum.

In asteroseismic modelling of massive stars there are degeneracies
between opacity, metallicity, age, mass, and overshooting \citep{A18}.
Having a tight constraint on stellar age obviously will remove at least
part of these degeneracies. Therefore, asteroseismic studies of runaway
pulsators 
with precise age determinations may become as important as the studies 
of pulsators in double-lined eclipsing binaries and can become vital in 
tracing the evolutionary history of such objects. HN~Aqr may not be the 
only star that can be studied that way (e.g., see Table~1 of 
\citealt{P09}).


The analysis of the first known $\beta$~Cep pulsator observed with {\it TESS} 
already demonstrates its potential for massive star asteroseismology. 
HN~Aqr, over the last 30 years believed to pulsate in a single 
frequency, exhibits at least 34 independent modes. This is the level of 
progress that space photometry of lower-mass stars has already achieved 
thanks to the Kepler mission; {\it TESS} is now opening the domain of massive 
stars for in-depth asteroseismology as well.\\






\acknowledgments

This paper includes data collected by the {\it TESS} mission. Funding for the 
{\it TESS} mission is provided by the NASA Explorer Program. Funding for the 
{\it TESS} Asteroseismic Science Operations Centre is provided by the Danish 
National Research Foundation (Grant agreement no.: DNRF106), ESA PRODEX 
(PEA 4000119301) and Stellar Astrophysics Centre (SAC) at Aarhus 
University. We thank the {\it TESS} team and staff and TASC/TASOC for their 
support of the present work. This work is also based on observations 
collected at the European Southern Observatory under ESO programme 
383.D-0909(A). Funding through the Polish NCN grants 
2015/18/A/ST9/00578, 2016/21/B/ST9/01126, 2015/17/B/ST9/02082 and 
2014/13/B/ST9/00902 is gratefully acknowledged. GMM acknowledges funding 
by the STFC consolidated grant ST/R000603/1. The research leading to
these results has received funding from the European Research Council
(ERC) under the European Union's Horizon 2020 research and innovation
programme (grant agreement no. 670519: MAMSIE). SS-D acknowledges
funding by the Spanish MCIU (projects AYA2015-68012-C2-1-P and SEV2015-0548)
and the Gobierno de Canarias (project ProID2017010115). GH thanks Daniel 
Heynderickx for supplying the photometry by \cite{WR88}, David Jones for 
help in retrieving archival data and Andrzej Baran for helpful comments 
on the manuscript.


\vspace{5mm}
\facilities{{\it TESS}, ESO VLT-UT2, UVES, SAAO}


\software{
MIARMA \citep{PGS15}, SigSpec \citep{R06}}



\clearpage

\appendix

\section{Online table}

\begin{deluxetable}{ccll}[htb!]
\tablecaption{Determinations of basic parameters of PHL~346.\label{tab:parameters}}
\tablecolumns{4}
\tablehead{
\colhead{$T_{\rm eff}$ (K)} & \colhead{log $g$} & \colhead{Ref.} & \colhead{Method}
}
\startdata
$22000 \pm 900$ & $3.4\pm0.2$ \tablenotemark{a} & \citet{KHB77} & Str\"omgren photometry\\
$21000 \pm 1500$ & $3.6 \pm 0.3$ \tablenotemark{b} & \citet{KL86} & Optical spectroscopy\\
$22600 \pm 1000$ & $3.6 \pm 0.2$ \tablenotemark{c} & \citet{K86} & Optical spectroscopy\\
$22900$ & $3.882$ \tablenotemark{d} & \citet{HWS94} & Walraven photometry\\
$22600$ & $3.890$ \tablenotemark{e} & \citet{HWS94} & Geneva photometry\\
$22300 \pm 1000$ & $3.7 \pm 0.2$ \tablenotemark{f} & \citet{R96} & Optical spectroscopy\\
$20700 \pm 1000$ & $3.58 \pm 0.10$ \tablenotemark{g} & \citet{RHM01} & Optical spectroscopy\\
$21500 \pm 900$ & 4.1 \tablenotemark{h} & \citet{ND05} & UV spectroscopy\\
$23800 \pm 900$ & $3.6\pm0.2$ \tablenotemark{j} & \citet{H11} & Str\"omgren photometry\\
$22290\pm450$ & $3.84\pm0.10$ \tablenotemark{k} & this work & Optical spectroscopy\\
$22400\pm300$ & $3.80\pm0.06$ \tablenotemark{l} & this work & Optical spectroscopy\\
\enddata
\tablenotetext{a}{$M_v=-3.3\pm0.4$, $E(b-y)=0.037$, calibration by \citet{NSW93}, $H_{\beta}$ from \citet{H11}}
\tablenotetext{b}{$v \sin i = 75\pm25$\,\kms, $E(b-y)=0.030$}
\tablenotetext{c}{Macroturbulence $\zeta = 12\pm3$\,\kms}
\tablenotetext{d}{$M_{bol}=-5.05$}
\tablenotetext{e}{$M_{bol}=-4.97$, calibration by \citet{NN90}}
\tablenotetext{f}{Macroturbulence $\zeta = 16\pm5$\,\kms}
\tablenotetext{g}{$v \sin i = 45$\,\kms}
\tablenotetext{h}{log $g$ derived from photometry, $[M/H]=0.21\pm0.09$, $E(B-V)=0.068\pm0.010$}
\tablenotetext{j}{$M_v=-3.3\pm0.4$, $E(b-y)=0.025$, calibration by \citet{NSW93}}
\tablenotetext{k}{$v \sin i=30.3\pm0.3$\,\kms, macroturbulence $\zeta = 18\pm2$\,\kms, microturbulence $\xi=8\pm1$\,\kms,  $[M/H]\approx0.3$, $RV=60\pm5$\,\kms (heliocentric)}
\tablenotetext{l}{$v \sin i=26\pm4$\,\kms, $\zeta=20\pm7$\,\kms, $\xi=14\pm1$\,\kms; abundances $C=8.78\pm0.07$, $N=8.20\pm0.05$, $O=8.90\pm0.08$, $Si=7.76\pm0.07$, and $Fe=7.66\pm0.07$}
\end{deluxetable}

\end{document}